\documentclass[preprint2]{aastex}
\usepackage{lscape}
\begin{document}

\title{Differential Radial Velocities and Stellar Parameters of Nearby 
Young Stars}

\author{Diane B. Paulson}
\affil{NASA's Goddard Space Flight Center, Code 693.0, Greenbelt MD 20771}
\email{diane.b.paulson@gsfc.nasa.gov}

\and

\author{Sylvana Yelda}
\affil{University of Michigan, 830 Dennison Building, Ann Arbor MI 48109}
\email{syelda@umich.edu}

\begin{abstract}
Radial velocity searches for substellar mass companions have focused primarily 
on stars older than 1~Gyr.
Increased levels of stellar activity in young stars hinders the detection of 
solar system analogs and therefore there has been a prejudice against 
inclusion of young stars in radial velocity surveys until recently.
Adaptive optics surveys of young stars have given us insight into the 
multiplicity of 
young stars but only for massive, distant companions. Understanding
the limit of the radial velocity technique, restricted to high-mass,
close-orbiting planets and brown dwarfs, we began a 
survey of young stars of various ages.
While the number of stars needed to carry out full analysis of the problems 
of planetary and brown dwarf population and evolution is large, 
the beginning of such a sample is included here.
We report on 61 young stars ranging in age from 
$\beta$ Pic association ($\sim$12 Myr) to the Ursa Majoris association 
($\sim$300 Myr). This initial search resulted in
no stars showing evidence for companions greater than $\sim$1-2 M$_{Jup}$ in
short period orbits at the 3$\sigma$-level. 
Additionally, we present derived stellar parameters,
as most have unpublished values. The chemical 
homogeneity of a cluster, and presumably of an association, may help to 
constrain true membership. As such, we present [Fe/H] abundances for the 
stars in our sample.
\end{abstract}


\section{Introduction}
The primary drive in extrasolar planet and brown dwarf detection
is to determine whether our solar system is unique and how our solar system 
formed and evolved into its present stablity. Currently,
radial velocity surveys probe nearby, solar-like stars for planets and
brown dwarfs. As new space-based
missions (e.g., Kepler, Gaia, SIM, TPF, and Darwin) are launched, the
question to be answered is whether habitable planets exist outside of our
solar system.
Precursor studies, like radial velocity and transit surveys, to these
missions must seek to understand the nearby
stellar systems that will be targets for these missions.

Focusing on solar-type stars, however, sidesteps the issues
regarding the formation and evolution of planets and brown dwarfs. To
understand these issues, we must survey young stars
for substellar mass companions. With the advance of adaptive optics and
interferometric imaging, various groups \citep[e.g., ][]{NeGuAl03, KaZuBe03,
MaMuHe05, Lowrance2005, ZuSoBe01}
have begun to do just this. These surveys are building reasonable
statistics of substellar mass companions around the youngest
stars. It is important to note that imaging surveys are most sensitive to
young, massive and distant
companions \citep{brandner2005}. As technology progresses, these instruments
will be able to reach lower masses with increased sensitivity and
with increased resolution, they will be able to probe the inner few AU
of nearby stars. The closest companions
(orbits of a few days - months) are unlikely to be detectable with direct
imaging anytime in the forseeable future.

The radial velocity method for detecting extrasolar planets 
is widely successful, having been used to detect over 150 planets to date
\citep{marcy2005}. Unfortunately, the technique
is subject to significant limitations. The method relies on
measuring the average line center for hundreds - thousands of stellar and
reference lines in order to achieve
the 2-3 m~s$^{-1}$ precision required to detect low mass planets around nearby
stars.  Anything that causes temporal changes in the line profiles,
including but not limited to strog magnetic fields, will adversely affect
this measurement and hence the radial velocity precision, as is discussed
by Saar \& Donahue (1997) and Hatzes (2002)\nocite{SaDo97, Ha02}.
\citet{Ha02}, \citet{SaDo97} and \citet{wright2005} discuss the details
of this issue and estimate the amplitude of this effect for various types
of stars. We must weigh the benefits of the radial velocity technique with
the pitfalls. The amplitude of the radial velocity perturbations induced
by a close orbiting high mass companion can be up to a few km~s$^{-1}$.
The intrinsic noise caused by stellar activity is typically less than
0.15~km~s$^{-1}$ for stars older than $\sim$50~Myr ($\S$4.1). Thus, high mass
short period companions can be detected in a straightforward
fashion using the radial
velocity method, and this is the focus of this paper\footnote{We also
would like to note the simultaneous efforts of \citet{esposito2005} in
carrying out a similar radial velocity search for extrasolar planets around
young stars.}.
It may be possible to remove the effects of stellar activity from
radial velocity measurements \citep[e.g., ][]{martinez2005,saar2003} in
order to search for lower mass companions, but we do not attempt such a feat.

\section{Sample}
For this study, stars were 
chosen in different age groups while a larger sample of ``other'' nearby young
stars was also included. The ages of stellar groups and associations range from 
the 12~Myr old $\beta$~Pic association \citep{ZuSoBe01} through the Ursa 
Majoris association, estimated to be $\sim$300~Myr 
\citep{soderblomandmayor1993}. 

The $\beta$ Pic association has gained much attention \citep[e.g., ][]{ZuSoBe01,
SoZuBe03, ortega04} in the past few years
because the members are at an interesting age, 
$\sim$12~Myr, roughly the age of 
circumstellar dust clearing.
This is the youngest group of stars in our survey and thus has 
significantly more intrinsic activity than any other star in this survey. For
these members, planetary mass companions are unlikely to be detected. However, 
binary and brown dwarf multiplicity is important for models of star
formation in stellar associations. It is for these reasons that 
imaging surveys of $\beta$~Pic members have been undertaken 
\citep[e.g., ][]{ZuSoBe01, NeGuAl03}. Abundance analysis of these stars
likely requires incorporation of non-LTE model atmospheres and such 
an analysis is not undertaken here.

IC~2391 is a cluster of age $\sim$30-50~Myr 
\citep{Mermilliod1981, barradoynavascues2004, pattenandsimon1996} at a 
distance of $\sim$150~pc \citep{Dodd2004, Stauffer1989}.
There are possibly as many as 125 members of IC~2391 \citep{Dodd2004}. We 
present analysis of only seven of these members. 
IC~2391 represents an excellent proving ground for studies of stellar 
activity \citep[e.g., ][]{marinoetal2005, simonandpatten1998} 
and angular
momentum evolution \citep[e.g., ][]{pattenandsimon1996,stauffer1997} 
because the stars are coeval, presumably of similar 
initial stellar composition and have ages between the youngest stars and the 
older clusters (such as the Pleiades). 
\citet{Randichetal2001} find log$\epsilon$(Fe)=7.49 for this cluster, placing
it just supersolar in metallicity (assuming log$\epsilon$(Fe)$_{\odot}$=7.45, 
Asplund, Grevesse \& Sauval 2005\nocite{asplund2005}). 

\citet{anosovaandorlov1991} first suggested the existence of the Castor moving 
group as an association of 15 stars. \citet{barradoynavascues1998} revisited 
this group of stars, adding additional ``probable'' members of the group, and 
determined an age of 200$\pm$100 Myr. Members were then added 
by \citet{montes2001}.  From these surveys, we
chose seven stars. Metallicity determination for this moving group has
not yet been explored in detail and our derived metallicities help 
constrain membership in this moving group.

The oldest association of stars we consider is the Ursa Majoris moving group.
Member stars are $\sim$300~Myr \citep{soderblom1993}. They are a
kinematically classified group of stars with, presumably, homogeneous
origins. Membership of this association has been carefully studied over the 
past several years (see, for example Soderblom \& Mayor 
1993\nocite{soderblomandmayor1993} and Montes et al. 2001).
Twelve stars are included in this study. These stars are also suited for more 
precise radial velocity
measurements for companion detection because by this age, the stars are less
active. The coarse radial velocity precision resulting from the use of
telluric lines dominates the measurement error. 
There are few nearby, bright stellar associations, so this becomes an 
important group for studies rearding chemical evolution.  The metallicity 
has been measured to be -0.09 \citep{BoFr90} and -0.05 \citep{KiSh05}.
\citet{KiSh05} further discuss the large scatter in abundances of
their sample of seven members and the possibility for initial chemical 
inhomogeneity. They also provide a nice tabulation of the abundances for
several other Ursa Majoris members from the literature.

Stars for our sample were chosen from two published surveys of nearby young 
stars \citet{montes2001} and \citet{ZuSoBe01}. For stars taken from
\citet{montes2001}, the star must have met the criterion that the
space velocity (U, V, W) agrees with a convergent point method for the
group and at least one of the
Eggen's kinematic criteria: peculiar or radial velocity, but
not necessarily both. 

We can also place possible constraints
on membership according to measured stellar abundance. Stellar clusters are 
formed from the
same material and as such should have the same initial stellar composition.
And, indeed, older clusters have remarkable consistency in [Fe/H],
\citep[e.g.][]{PaSnCo03}. Large deviations in stellar composition may indicate
either pollution of the stellar photosphere or protostellar nebulae of unequal
composition. Because of the ambiguity, we can only offer a new label of
``possible non-member''. Discrepancies of stellar abundance with respect to
cluster or association mean have been pointed out in each of the Results
sections.

A further restriction was placed on the
rotational velocity of the star. The higher mass stars (earlier than F8) have
higher rotation rates which broaden the stellar lines
and reduce the velocity precision. Typically
radial velocity surveys which employ the I$_{2}$ cell place a 10-15~km~s$^{-1}$
cutoff for maximum stellar $v$sin$i$, giving preference to
stars with much lower $v$sin$i$. Because we are not attempting to achieve
3~m~s$^{-1}$ precision, this constraint was relaxed to $\sim$20~km~s$^{-1}$.

\section{Observations}
The observing program was designed to maximize efficiency and this project 
lends itself well to quick detection of high mass, close-in companions. The 
goal was to survey
about two dozen stars during a 3-4 night run by taking 2-4 spectra of
each star each night. During the next run, stars from the previous runs that 
showed
large amplitude variability over a few day's time were reobserved. In this 
way, more than a dozen stars were eliminated as having no close-in, high-mass 
companion and did not need to be reobserved. We surveyed a total of 61 stars. 

Of the nights granted for this project on the Magellan telescopes, data were 
collected on the nights listed in Table 1. The MIKE spectrograph \citep{BeSh03}
was used with the 0.35" slit, yielding a resolving power of $\sim$54,000 on
the red chip and $\sim$70,000 on the blue chip. The red grating angle was
offset twice during the course of these observations, and in Table 1 
we list the wavelength ranges for each observing run.
The blue chip went unchanged throughout the course of this investigation. No
noticeable offset was observed between shifts of the spectrograph to within
errors of our technique. Observations were only recorded when seeing was sub-2".
Exposures were constrained to 30 minutes or less to maximize velocity 
precision, with typical exposures ranging from 2-15~minutes.

Spectra were reduced using standard IRAF\footnote{IRAF is distributed by the National Optical Astronomy Observatories,
    which are operated by the Association of Universities for Research
    in Astronomy, Inc., under cooperative agreement with the National
    Science Foundation.} packages. After flat fielding and removal of the 
bias with the ccdproc routine, we used 
a suite of IRAF routines, mtools, for use in specifically handling MIKE 
spectra. The mtools routines were 
developed by J. Baldwin and are publicly available at the Las Campanas 
Observatory website (www.lco.cl). MIKE spectra have a tilt of the slit with 
respect to the orders \citep{BeSh03}. The mtools routines
calculate the varying slope of the tilt using ThAr calibration spectra
along the orders and across each chip. It is repeated for each of the ThAr 
spectra taken throughout the night.
The tilt is then removed from all spectra during the
extraction of the spectral orders.
Once the spectra have been corrected for the tilt,
a wavelength solution is applied using the IRAF package ecid. 

In lieu of an I$_{2}$ cell, we took advantage of the telluric oxygen band
at 6900\AA\ for our velocity reference. Our
philosophy in not using the I$_{2}$ cell
was that the intrinsic noise of these objects will exceed the few m~s$^{-1}$
precision one can achieve with the cell. 
Longer term projects, investigating either the intrinsic and variable
activity of the stars or the presence of longer period high mass 
companions, would benefit from the use of an I$_{2}$ cell. 

\section{Analysis}
\subsection{Radial Velocities and Errors}
The MIKE spectrograph
is not fiber-fed and thus the measurement of radial velocities to this
precision is hindered by guiding errors, however this is not easily corrected
for. Variable positioning of the stellar image on the slit illuminates the 
spectrograph non-uniformly. This variable illumination causes inherent shifts 
of the stellar absorption features.  The use of telluric 
features as a velocity reference only corrects for part of this error, as the 
telluric features are themselves broad. Instrumental errors are also introduced 
by temperature fluctuations in the spectrograph, as it is not in a temperature controlled environment. These are removed by taking frequent ThAr
spectra for contemporaneous wavelength solutions. As a check that this
was necessary, we measured the relative velocity shifts of the ThAr spectra
throughout the night.
The maximum variation throughout the course of one observing night was
0.7~km~s$^{-1}$. Thus, ThAr
spectra were taken between 10 and 20 times each night for wavelength
calibration. The wavelength solution nearest in time to the stellar 
spectrum was applied.

Radial velocities are measured via cross-correlation of each observation
with a ``template'' spectrum of that
same star using the IRAF package $fxcor$. The template spectrum was
chosen to be the spectrum with the
highest signal-to-noise. We subtracted the differential velocity shift
of the telluric lines from those of the stellar lines to obtain the final
radial velocity for that observation in the heliocentric frame of reference
(provided within $fxcor$). The value $A_{obs}$ (in
Tables 2, 3, 5, 7 and 9) is the 1-$\sigma$ rms of all velocity measurements for
each star. Each observation also includes an associated error.
This error is determined by adding in quadrature the
cross correlation error derived in $fxcor$ for the stellar lines with that of
the telluric features.
The error listed for each star, $\sigma_{\rm error}$ in the Tables, as well
as in Figure 1, is the rms of the errors for all observations of a given star.

Under good seeing conditions ($\sim$0.5") and on bright (V$\sim$8), 
slowly
rotating stars we were able to obtain velocity precision of 0.01~km~s$^{-1}$
for an individual observation, with an rms of 0.02-0.03~km~s$^{-1}$.
Two stars known to not harbor hot Jupiters were observed: $\tau$~Ceti and 
HD~99109. We chose these to look for 
systematics during the course of observations, though no systematics were 
found at our detection threshold. These two stars also provide 
confirmation of our above quoted errors. 
$\tau$~Ceti was observed 11 times over 7 nights and exhibited velocity
variations A$_{obs}$=0.03~km~s$^{-1}$ with 
$\sigma_{\rm error}$=0.04~km~s$^{-1}$. And, HD~99109 which was observed 
16 times over 10 nights has 
A$_{obs}$=0.04~km~s$^{-1}$ and $\sigma_{\rm error}$=0.04~km~s$^{-1}$.
Our somewhat crude velocity precision is sufficient for 
detection of high mass, close-in planets and brown dwarfs
as the velocity variations will exceed the combination of the 
intrinsic noise caused by stellar activity
and the instrumental error (which includes the uncertainties arising from 
use of telluric features).

We adopt the empirical relationship of \citet{SaDo97}, using our measured 
rotational velocities and Hipparcos photometry, to compare observations
with predicted radial velocity amplitudes induced by the 
rotational modulation of starspots.
According to the Saar \& Donahue relationship, G-type stars of Hyades age 
($\sim$650 Myr) will have a maximum predicted radial velocity amplitude of
$\sim$30-40~m~s$^{-1}$, while G-type stars of $\beta$~Pic age ($\sim$12 Myr)
can exhibit radial velocity variations $\sim$0.7~km~s$^{-1}$. It is apparent 
that the only detectable companions around the youngest stars will be close-in 
brown dwarfs and stellar binary companions. Any information on the 
presence of companions (stellar or otherwise) around these young stars is 
useful for understanding the dynamics of post-T Tauri systems.
For the calculation of predicted maximum amplitudes, 
$A_{pred}$=0.6$f_{S}^{0.9}$$v$sin$i$ (where $f_{S}$=$\sim$0.4$\Delta$V is 
given in 
percent) we make use of our measured $v$sin$i$ ($\S$4.3), when available
and the Hipparcos photometric database 
\citep{ESA}, unless otherwise noted. The adopted and measured values are 
listed in Tables 2, 3,
5, 7 and 9. Hipparcos measures magnitudes in a slightly different 
bandpass as V, but the amplitude of the variability is similar. 
From the database, we take 
as the amplitude the difference between the 95\% and the 5\% photometric 
levels. We list these values as $\Delta$V$_H$.  Observed versus 
predicted radial velocity amplitudes, including velocity errors, are shown 
in Figure 2. A 1:1 line is drawn for reference
as well as a (dashed) line representing the mean error for all observations.
Two stars have arrows indicating the questionable Hipparcos photometry. The 
end points of these arrows are the values obtained if we assume photometry 
taken from ground-based surveys.

\subsection{Limits on Detectable Companions}
Using the methods of \citet[][ hereafter NA98]{NeAn98} as we did in 
\citet{PaCoHa04}, we 
determine limits on companions detectable given the data quantity and
quality for each star (as listed in the respective tables). The error 
we assume is the predicted velocity caused by stellar activity, $A_{pred}$,
added in quadrature with the mean observational error, $\sigma_{\rm error}$.
The value  $M_{3\sigma}$ 
is the lower mass limit of companions ($>$1M$_{Jup}$) that can be ruled
out at the 3-$\sigma$ level in orbits of 6 or fewer days.
The number of independent 
frequencies sampled is given in Eq. 14 of NA98. We have 
assumed even sampling, but as our data is not uniformly sampled, it is 
possible to search much higher frequencies than then Nyquist frequency. 
The velocity that will be exceeded in a given frequency range, $K$, is 
dependent on 
the square root of $ln$(1/$N$), where $N$ is the number of 
independent frequencies in a given frequency range (see Eq. 15 in 
NA98). Thus, a change in the number of independent frequencies by a factor of 
2 only results in a fractional change of $K$. And, as such, our assumption 
of evenly sampled data does not significantly affect the results of this 
analysis.
Upon inspection of the table, the most massive planets and brown dwarfs are
ruled out for all stars. For most stars, planets as small as 1~M$_{Jup}$
can also be ruled out. The mass limit of companions around these stars can be 
decreased with an increased number of observations as well as more precise 
radial velocities for the less active stars. However, the mass limits for the 
youngest stars (such as the $\beta$~Pic members) remain limited by the 
intrinsic activity of the star.

\subsection{Stellar Parameters and Abundance Analysis}
In order to measure rotational velocities, $v$sin$i$, for each star, we
employ the curve of growth method to determine stellar parameters. These 
parameters are then used to model a small spectral section to 
extract line broadening resulting from stellar rotation. 

For the majority of stars in this sample, we have derived the stellar
parameters and metallicity using the spectral synthesis code MOOG
\citep{Sn73}, in the same fashion as we did in \citet{PaSnCo03}. 
It has been recently shown that cooler stars exhibit an ionization 
imbalance which increases as the stellar temperature decreases 
\citep[e.g.,][]{allendeprieto2004, yong2004}. Allende Prieto et al. (2004)
suggest that this, sometimes severe, departure from LTE stems from an 
increase in stellar activity (in cooler stars relative to warmer stars 
with shallower convective regions). If this is indeed the 
root cause of these discrepancies, then the increase in activity as a result 
of stellar youth would also imply the requisite use of non-LTE stellar 
atmopsheres.  We did not attempt measurement of stellar parameters for
the $\beta$ Pic members and stars cooler than K2 because of this issue. We
also note that for some elements, the use of LTE models may result in 
incorrect abundances for all stars in this survey because of their youth.

We use stellar atmosphere models based on the
(LTE) 1995 version of the ATLAS9 code \citep{Castelli1997}, and while newer 
non-LTE models have recently become available \citep[e.g.][]{hauschildt1999}, 
they require too great of 
computational time to obtain a reasonably large grid of atmospheres. 
We have thus adopted $v$sin$i$ values from the literature (references 
are noted in the tables) for those stars for which we have not derived 
atmospheric parameters. 

Within IRAF, EWs are measured by fitting Gaussians to Fe~I and Fe~II lines 
(see Table 1 in Paulson, 
Sneden \& Cochran 2003 for the linelist used).
The EWs are input into MOOG drivers along 
with the individual line parameters to back out the effective temperature, 
$T_{\rm eff}$, surface gravity, log$g$, microturbulence, $\xi$, 
metallicity, [Fe/H]
and finally, the rotational velocity, $v$sin$i$. The $v$sin$i$s are measured
by spectral synthesis after determining the appropriate stellar
atomosphere model, via the
curve of growth routine ($lin$), by employing the $synth$ feature of MOOG. 
Our grid of models dictates the errors on $T_{\rm eff}$, log$g$ and $\xi$ to 
be 50~K, 0.1~cm~s$^{-2}$ and 0.2~km~s$^{-1}$, respectively.
In order to determine [Fe/H], we require that the $T_{\rm eff}$  be 
independent of excitation potential for all Fe~I lines, 
and $\xi$ must be independent of line strength. Upon fitting these two 
parameters, 
the surface gravity is determined by requiring ionization equilibrium between
Fe~I and Fe~II. These three parameters are iterated upon until these 
requirements are all met. An overall [Fe/H] abundance is then calculated based
on a mean of the the individual line abundances. Using this model, we 
synthesize a spectral 
region containing five Fe~I lines of varying excitation potential. The 
only parameter that is varied is the rotational velocity. 
We do not deconvolve the macroturbulence from the rotational
velocities, so the $v$sin$i$ values listed in Tables 2, 3, 5, 7, and 9 
include the effects of macroturbulence.  Observing
the solar spectrum and following the exact same analysis as we do for all stars 
in this paper enables us to compare our values to solar in a straightforward
manner. This approach removes the possible systematics that may arise from
this instrument, and thus we present absolute [Fe/H] (with errors 
$\sim$0.08 dex) values independent of 
the value of log$\epsilon$(Fe)$_{\odot}$. 
Our solar spectrum was taken of the asteroid Iris, as it 
closely resembles starlight through the combination of the telescope and 
the instrument. The final stellar parameters are listed in Tables 4, 6, 8, 
and 10.

\section{Results}
\subsection{$\beta$ Pic Moving Group}
We included the $\beta$~Pic members as a demonstration of the limiting 
factor of radial velocity measurements on the youngest stars. As seen in Table 
2, the radial velocity amplitudes of these stars are quite large, 
0.25-0.61~km~s$^{-1}$. The observed amplitudes are within the predicted values 
except for three stars. GJ~803, GJ~3305 and HIP~88399 have 
$\Delta$V$_{H}$ magnitudes 
well below that of other members. For GJ~803, ground based photometry produced 
$\Delta$V = 0.22 (Custipoto 1998). This value would bring $A_{pred}$ to 
0.41~km~s$^{-1}$. This rectifies the discrepancy between predicted and observed
amplitudes. If we assume $\Delta$V$_{H}$ to be ~0.2 for HIP~88399, in line 
with other member's $\Delta$V$_{H}$, then $A_{obs}$ becomes 
0.85~km~s$^{-1}$, well in agreement with observation.
GJ~3305 also has a remarkably low photometric amplitude. Because of lack of 
$v$sin$i$ information, a comparison of $A_{pred}$ with $A_{obs}$ is not 
possible. If the low photometric amplitudes for GJ~3305 and HIP~88399 are
indeed real, this would suggest that the stellar inclinations  were almost
pole-on. But, at least for the case of HIP~88399, the rotational velocity 
indicates otherwise. So, it is probable that the photometry is not accurate.
Four of the stars in our survey: GJ~803, HIP~23309, HIP~29964, and HIP~76629
have been surveyed with adaptive optics by 
\citet{NeGuAl03} and were shown to harbor no distant binary companions. Thus, 
we can rule out brown dwarf companions and stellar companions in very close 
obrits as well as in several AU orbits. 

\subsection{IC~2391}
Solar analogs in the IC~2391 
cluster will have just arrived on the main sequence and are thus an important 
sample of stars for multiplicity studies. The X-ray levels of stars in this 
cluster confirm the age of 30 Myr \citep{marinoetal2005} and the rotation 
of members is also consistent with that age.
Despite its youth, IC~2391 has fairly low activity levels as measured by
photometric variability. This is likely caused by somewhat uniform coverage of
starspots. Thus, the radial velocity variations 
caused by photospheric activity are also lessened. 
As a result, there seems to be a division in the accessibility
of planet and brown dwarf detection via the radial velocity method between 
the ages of $\beta$~Pic and IC~2391. Members of IC~2391 are good candidates for
longer term follow up (i.e. for substelar mass companions in orbits of a few 
weeks - months). The observed radial velocities (Table 3)
agree to within error of the expected amplitude, and companions of 
mass $>$1-2~M$_{Jup}$ with orbital periods less than 6 days can be ruled out.

The stellar parameters for observed IC~2391 members are listed in Table 4.
The average [Fe/H] of IC~2391 
is -0.01 $\pm$ 0.12. We note, however, that BD+01~2063 is 
much lower in abundance than the remaining cluster members with an [Fe/H] of
-0.24. If this star is truly a member of IC~2391, as 
indicated by the positive results of the two membership tests 
\citet{montes2001} performed,
then further studies into the depletion of metals in this star are warranted.
However, it is more likely that this star is an interloper whose stellar history
is vastly different than the surrounding cluster. Removing BD+01~2063 
from the group results in a mean [Fe/H] = 0.03.

\subsection{Castor Moving Group}
The members of the Castor moving group we surveyed also show remarkable 
photometric stability and thus stability in the radial velocity measurements. 
Despite their youth, stars in the Castor group, like IC~2391, are good 
candidates for more precise and longer-term radial velocity searches.
None of the members of the Castor moving group show radial velocity variability
indicative of short-period companions (Table 5) and companions as low as 1~M$_{Jup}$ with short-periods are not detected. We measure a mean 
[Fe/H] of Castor members of 0.00$\pm$0.04. While our sample (5 stars) is small, 
there is very little scatter in the abundances. 

\subsection{Ursa Majoris Moving Group}
As for the previous groups, there are no stars whose $A_{obs}$ significantly
exceeds $A_{pred}$. As a result, companions with mass $>$1-2~M$_{Jup}$ in close-orbits are ruled out.
The previously measured [Fe/H] of this association is between -0.05 and -0.09
($\S$2). 
We determine an [Fe/H] of 0.06$\pm$0.11.
We attribute the difference between our measurement and previous measurements 
to the samples of stars chosen for each survey.
There is a scatter in the abundances 
of ``members'' of 0.42~dex ($\sigma$ = 0.11~dex). We assume (perhaps
incorrectly?) that these stars were formed from homogeneous material.
The spread of [Fe/H] of these stars is not easily explained 
for a common group, i.e. scatter in the Hyades is $\pm$0.04~dex 
\citep[][ {\rm [Fe/H]} = 0.13]{PaSnCo03} and in the
younger Pleiades $\pm$0.05 \citep[][ {\rm [Fe/H]} = 0.06]{kingetal2000}. 
Four of the stars we include have abundances which are $>$0.10~dex higher
(or lower) than the cluster median- BD+19~2531, HD~64942, HD~88654 and 
HD~131156A. The
membership of these stars to the Ursa Majoris association, their 
overall metallicities and the overall abundance variations throughout the 
entire association should be explored further. 

\subsection{Other Nearby Young Stars}
Some of the stars we include here have confirmed binary companions as
noted in the literature, though our data indicate
that none of them have high mass short-period orbiting companions (stellar or
substellar). Those stellar companions cited in the literature include only 
distant companions whose radial velocity effects would be immeasurable in 
our data.
\citet{balega2002} found a companion to HD~15013 with a separation of
127~mas. \citet{gaidos1998} indicate HD~96064 has a probable M dwarf 
companion at 270~AU. \citet{Lowrance2005} 
find that HD~82443 has a binary companion at a distance of 6.86". The only
star with large $A_{obs}$ is
HD~82558 which has a $v$sin$i$ =28.0~km~s$^{-1}$ \citep{kovari2004}.
The $v$sin$i$ is reflected in $A_{pred}$ and is much greater than $A_{obs}$.
None of these stars has a companion with mass $>$1-2~M$_{Jup}$ in a close orbit.

\section{Discussion}
We present in this paper an initial search of young stars at different ages 
for substellar mass companions as well as a census of
stellar parameters and metallicity. While no companions are detected in 
this sample, we list mass limits for rejected planets and brown dwarfs.
Stars of age $\sim$IC~2391 and older are suitable for 
long-term radial velocity surveys for high-mass, short-period 
($\sim$few months) companions based on their relatively stable activity 
levels. For younger stars, like $\beta$~Pic 
members, we are limited to searches for short-period brown dwarf 
companions. For these stars, the velocity technique employing telluric lines 
as a velocity reference is sufficient. The sub-5~m~s$^{-1}$ precision gained
through use of an I$_{2}$ cell provides little improvement to a star whose 
intrinsic activity levels are greater than 0.05~km~s$^{-1}$. Thus, for
longer-term follow-up of stars in our paper, velocity precision of 
5-10~m~s$^{-1}$ is more than adequate. 

With sufficient telescope allocation, it is possible to look 
for longer period planetary mass companions using the radial velocity 
technique. In order to do so, one could average the 
velocity over the course of a rotation cycle, as the velocity amplitude will be
(or at least come close to) zero. Of course, small variations in total spot 
coverage may cause ``zero point" offsets, but this technique should be able to 
allow for detection of companions in orbits of a few weeks-months. 
Unfortunately, 
this type of project would be considerably less efficient than the current 
radial velocity planet searches by lowering the number of stars that 
could be surveyed. 

The lack of ``hot Jupiters" in this sample is not surprising. Only 0.8\% of
stars have such objects within 0.1~AU of the parent star \citep{marcy2005b}.
Our sample is
considerably smaller than what would be required to have a meaningful
statistic. However, we can state that the incidence of such objects around
young stars is less than 1.5\% and likely much smaller, in agreement with the
0.8\% incidence of companions around old stars. To build better
statistics, a comprehensive and dedicated survey similar to the N2K project 
\citep[e.g., ][]{fischer2005} should be carried out for the youngest 
nearby stars. 

We further recommend combining radial velocity surveys for close-in companions 
with imaging surveys for distant companions.
The radial velocity technique is limited to
detection of only high-mass close-in companions, whereas AO and other imaging
techniques are only sensitive to more distant, young, high-mass companions. 
Combining these types of surveys on the same set of young stars 
is the best way to begin to understand 
the evolution and formation of planets and brown dwarfs at these young and 
intermediate ages. 

\acknowledgments
This research made extensive use of the SIMBAD astronomical database, operated
at the Centre de Donnees Astronomiques de Strasbourg, France and NASA's 
Astrophysics Data System and the arrow.pro procedure written and copyrighted
by Johns Hopkins University/Applied Physics Laboratory.
Data presented herein were obtained at the Magellan Telescopes at the Las
Campanas Observatory operated by the Observatories of the Carnegie 
Institution of Washington, the University of Arizona, University of Michigan,
Harvard University and the Massachusettes Institute of Technology.
DBP wishes to thank the 
Department of Astronomy at the University of Michigan for granting the 
telescope time needed for this project and the
telescope operators and instrument scientists for making sure that our
observations ran smoothly.
DBP wishes to thank W. Cochran for his continuing support and useful
discussions on planet searches.  DBP is currently a National Research Council
fellow working at NASA's Goddard Space Flight Center.
DBP would also like to acknowledge the
financial support from E. Bergin (NASA grant NAG5-9227
and NSF grant AST-9808980) during a portion of this research.
 

\begin{figure}
\plotone{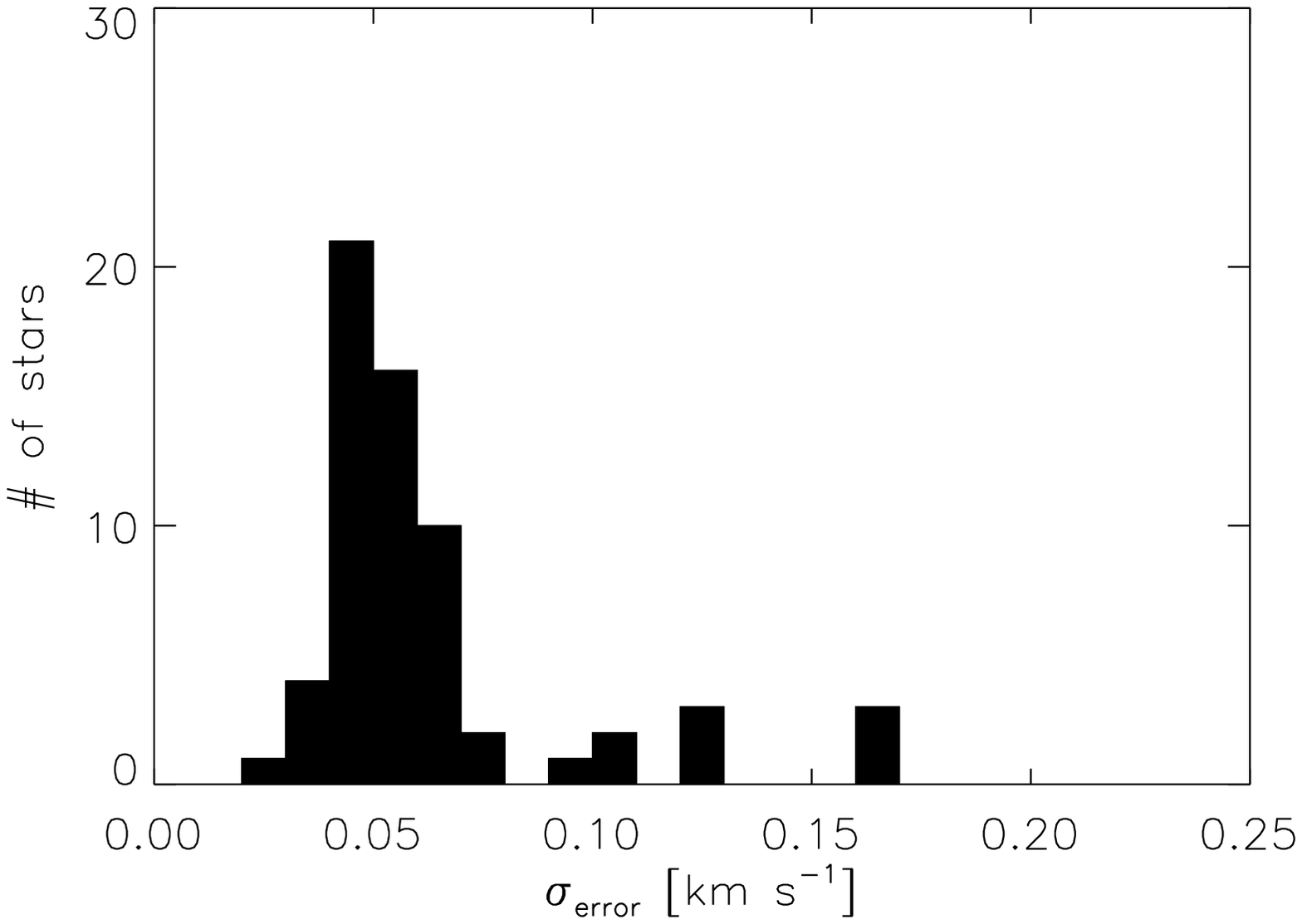}
\caption{A histogram of the mean error for each star. Each component of this
histogram is the rms of the errors from all observations of a given star.}
\end{figure}

\begin{figure}
\plotone{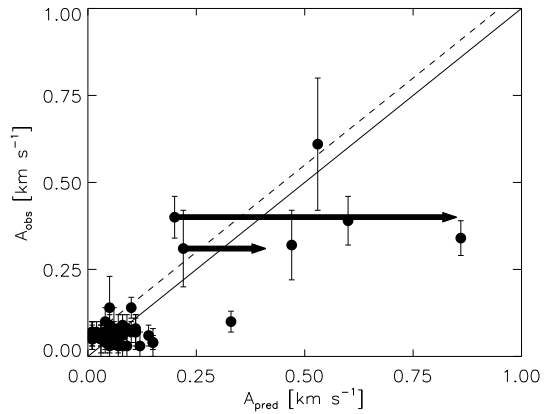}
\caption{Observed versus predicted radial velocity amplitudes. Arrows 
shown correspond to the two stars in $\beta$~Pic which have questionable 
photometry, as discussed in the text. A ($solid$) 1:1 line is drawn along 
with a ($dashed$)
line corresponding to the mean error of all data.}
\end{figure}
 
\begin{deluxetable}{lcccc}
\tabletypesize{\scriptsize}
\tablecaption{Observing Log}
\tablewidth{0pt}
\tablehead{
\colhead{Dates} & \colhead{MIKE Red Range [\AA]} & \colhead{MIKE Blue 
Range [\AA]} }
\startdata
October 2 - 5, 2003 	& 4500 - 7100 & 3700 - 4800\\
December 9 - 10, 2003	& 4500 - 7100 & 3700 - 4800\\
March 10 - 13, 2004	& 4750 - 8500 & 3700 - 4800\\
April 2 - 5, 2004	& 4750 - 8500 & 3700 - 4800\\
June 1 - 3, 2004 	& 4900 - 8900 & 3700 - 4800\\
July 15 - 19, 2004 	& 4900 - 8900 & 3700 - 4800\\
March 20 - 21, 2005	& 4900 - 8900 & 3700 - 4800\\
\enddata
\end{deluxetable}

\begin{deluxetable}{lcccccccccc}
\tabletypesize{\scriptsize}
\tablecaption{$\beta$ Pic Moving Group Velocities}
\tablewidth{0pt}
\tablehead{
\colhead{Star} & \colhead{\# nights obs}&
\colhead{\# obs}&
\colhead{$v$sin$i$} & 
\colhead{$\Delta$V$_{H}$} &
\colhead{$A_{pred}$} & 
\colhead{$A_{obs}$}  &
\colhead{$\sigma_{\rm error}$} &
\colhead{M$_{3\sigma}$} & 
\colhead{Notes} \\ 
\colhead{} & 
 &  & 
[km s$^{-1}$] & 
[mag] &
[km s$^{-1}$] & 
[km s$^{-1}$] & 
[km s$^{-1}$] & [M$_{Jup}$]  &
 }
\startdata
GJ 803    &  8  & 13 & (9)     & 0.11 & 0.22    & 0.31 & 0.16 & \nodata &  1,2 \\
GJ 3305   &  5  & 11 & \nodata & 0.02 & \nodata & 0.25 & 0.12 & \nodata &  1 \\
HIP 23309 &  13 & 23 & (11)    & 0.20 & 0.47    & 0.32 & 0.16 & 9 &  1 \\
HIP 29964 &  16 & 25 & (13)    & 0.22 & 0.60    & 0.39 & 0.12 & 11 &  1  \\
HIP 76629 &  14 & 35 & (11)    & 0.23 & 0.53    & 0.61 & 0.36 & 8 &  1 \\
HIP 88399 &  7  & 16 & (20)    & 0.04 & 0.20    & 0.40 & 0.10 & \nodata &  1,3 \\
\enddata
\\
(1) Adopted $v$sin$i$ values from \citet{ZuSoBe01}.
(2) \citet{Cutispoto1998} give $\Delta$V=0.22. Adopting this 
value would give a predicted amplitude of 0.414.
(3) If we adopt an average photometric 
amplitude of 0.2 (typical for this association) instead, the predicted 
amplitude becomes 0.845 km~s$^{-1}$.
\end{deluxetable}

\begin{deluxetable}{lccccccccc}
\tabletypesize{\scriptsize}
\tablecaption{IC~2391 Radial Velocities}
\tablewidth{0pt}
\tablehead{
\colhead{Star} &\colhead{\# nights obs}&
\colhead{\# obs}&
\colhead{$v$sin$i$} &
\colhead{$\Delta$V$_{H}$} &
\colhead{$A_{pred}$} &
\colhead{$A_{obs}$}  &
\colhead{$\sigma_{\rm error}$} &\colhead{M$_{3\sigma}$} & 
\colhead{Notes}\\ 
\colhead{} & \colhead{} & & 
[km s$^{-1}$] &
[mag] &
[km s$^{-1}$] &
[km s$^{-1}$] &
[km s$^{-1}$] & [M$_{Jup}$] &  
\colhead{}}
\startdata
BD+01 2063 &  7  & 14 & 4.0    & 0.05 & 0.05 & 0.03 & 0.02 & 1 &  \\
HD 111813  &  3  &  4 & 3.8    & 0.08 & 0.07 & 0.05 & 0.06 & 1 &  \\
HD 118100  &  4  & 10 & (14.0) & 0.15 & 0.33 & 0.10 & 0.04 & 2 & 1 \\
HD 120352  &  4  &  9 & 3.2    & 0.05 & 0.04 & 0.06 & 0.04 & 1 &  \\
HD 140913  &  1  &  3 & 9.0    & 0.07 & 0.15 & 0.04 & 0.03 & 2 &  \\
HD 142072  &  3  &  8 &6.1     & 0.05 & 0.07 & 0.08 & 0.06 & 1 &  \\
HD 157750  &  4  &  8 &3.4     & 0.05 & 0.04 & 0.07 & 0.05 & 1 &  \\
HD 209779  &  2  &  7 &6.8     & 0.04 & 0.07 & 0.03 & 0.06 & 1 &  \\
\enddata
\\
(1) $v$sin$i$ from \citet{cutispoto2002}
\end{deluxetable}

\begin{deluxetable}{lcccc}
\tabletypesize{\scriptsize}
\tablecaption{IC~2391 Stellar Parameters}
\tablewidth{0pt}
\tablehead{
\colhead{Star} & \colhead{$T_{\rm eff}$} & \colhead{log$g$} & \colhead{$\xi$} &
\colhead{[Fe/H]}\\ 
\colhead{}
& [K] & [cm s$^{-2}$] & [km s$^{-1}$] & }
\startdata
BD+01 2063 & 5350 & 4.5 & 0.8 & -0.24 \\
HD 111813 & 5000 & 4.6 & 0.8 & -0.08 \\
HD 118100 & \nodata & \nodata & \nodata & \nodata\\
HD 120352 & 5500 & 4.5 & 0.8 & -0.03 \\
HD 140913 & 6050 & 4.4 & 0.8 & 0.06 \\
HD 142072 & 5900 & 4.4 & 0.8 & 0.11 \\
HD 157750 & 5850 & 4.4 & 0.5 & 0.10 \\
HD 209779 & 5850 & 4.4 & 0.5 & 0.04 \\
\enddata
\end{deluxetable}

\begin{deluxetable}{lccccccccc}
\tabletypesize{\scriptsize}
\tablecaption{Castor Moving Group Radial Velocities}
\tablewidth{0pt}
\tablehead{
\colhead{Star} &\colhead{\# nights obs}&
\colhead{\# obs}&
\colhead{$v$sin$i$} &
\colhead{$\Delta$V$_{H}$} &
\colhead{$A_{pred}$} &
\colhead{$A_{obs}$}  &
\colhead{$\sigma_{\rm error}$} &\colhead{M$_{3\sigma}$} & 
\colhead{Notes}\\
\colhead{}
 & & & 
[km s$^{-1}$] &
[mag] &
[km s$^{-1}$] &
[km s$^{-1}$] &
[km s$^{-1}$] & [M$_{Jup}$] &  }
\startdata
BD+24 2700 & 4  &   8 & 2.0    & 0.06 & 0.03   & 0.06 & 0.05 & 1  &\\
HD 41842   & 10 &  22 & 3.0    & 0.08 & 0.06   & 0.05 & 0.04 & 1  &\\
HD 77825   & 5  &  10 & 2.5    & 0.07 & 0.04   & 0.07 & 0.04 & 1  &\\
HD 94765   & 6  &  11 & 2.0    & 0.07 & 0.03   & 0.05 & 0.03 & 1  &\\
HD 103720  & 5  &   9 & \nodata& 0.08 & \nodata& 0.07 & 0.04 & 1  &\\
HD 181321  & 3  &   8 & 14.0   & 0.04 & 0.14   & 0.06 & 0.04 & 2  &\\
HD 216803  & 3  &   7 & (3.0)  & 0.04 & 0.03   & 0.05 & 0.05 & 1  &1\\
\enddata
\\
(1) $v$sin$i$ from \citet{nordstrom2004}
\end{deluxetable} 

\begin{deluxetable}{lcccc}
\tabletypesize{\scriptsize}
\tablecaption{Castor Moving Group Stellar Parameters}
\tablewidth{0pt}
\tablehead{
\colhead{Star} & \colhead{$T_{\rm eff}$} & \colhead{log$g$} & \colhead{$\xi$} &
\colhead{[Fe/H]}\\
\colhead{} &  [K] & [cm s$^{-2}$] & [km s$^{-1}$] & \colhead{}}
\startdata
BD+24 2700 & 5100 & 4.6 & 0.8 & 0.02 \\
HD 41842   & 5150 & 4.5 & 1.0 & -0.03 \\
HD 77825   & 5100 & 4.5 & 1.0 & -0.05 \\
HD 94765   & 5050 & 4.5 & 0.8 & 0.03 \\
HD 103720  & \nodata & \nodata & \nodata & \nodata\\
HD 181321  & 6000 & 4.5 & 1.0 & 0.04 \\
HD 216803  & \nodata & \nodata & \nodata & \nodata\\
\enddata
\end{deluxetable}

\begin{deluxetable}{lcccccccc}
\tabletypesize{\scriptsize}
\tablecaption{Ursa Majoris Moving Group Radial Velocities}
\tablewidth{0pt}
\tablehead{
\colhead{Star} &\colhead{\# nights obs}&
\colhead{\# obs}&
\colhead{$v$sin$i$} &
\colhead{$\Delta$V$_{H}$} &
\colhead{$A_{pred}$} &
\colhead{$A_{obs}$}  &
\colhead{$\sigma_{\rm error}$} &\colhead{M$_{3\sigma}$} \\
\colhead{}
 & & & 
[km s$^{-1}$] &
[mag] &
[km s$^{-1}$] &
[km s$^{-1}$] &
[km s$^{-1}$] & [M$_{Jup}$] }
\startdata
BD+19 2531 & 6  &   8 & 1.5 &  0.02 &  0.01 & 0.06 & 0.04 & 1  \\
HD 26913   & 13 &  23 & 8.5 &  0.04 &  0.08 & 0.09 & 0.04 & 2  \\
HD 38392   & 9  &  16 & 1.0 &  0.05 &  0.01 & 0.07 & 0.04 & 2  \\
HD 41593   & 4  &   6 & 3.8 &  0.10 &  0.09 & 0.03 & 0.06 & 1  \\
HD 60491   & 5  &  13 & 5.4 &  0.06 &  0.08 & 0.06 & 0.04 & 1  \\
HD 61606   & 11 &  25 & 2.2 &  0.05 &  0.03 & 0.05 & 0.07 & 1  \\
HD 64942   & 12 &  28 & 8.5 &  0.05 &  0.10 & 0.07 & 0.05 & 2  \\
HD 81659   & 8  &  19 & 1.0 &  0.06 &  0.01 & 0.05 & 0.03 & 1  \\
HD 88654   & 6  &  17 &\nodata&0.03 &\nodata& 0.04 & 0.04 & \nodata  \\
HD 98712   & 11 &  21 & 4.5 &  0.04 &  0.05 & 0.09 & 0.05 & 2  \\
HD 131156A & 4  &   9 & 3.7 &  0.04 &  0.04 & 0.10 & 0.05 & 1  \\
HD 165185  & 4  &   8 & 7.7 &  0.05 &  0.09 & 0.08 & 0.06 & 1  \\
\enddata
\end{deluxetable}

\begin{deluxetable}{lcccc}
\tabletypesize{\scriptsize}
\tablecaption{Ursa Majoris Moving Group Stellar Parameters}
\tablewidth{0pt}
\tablehead{
\colhead{Star} & \colhead{$T_{\rm eff}$} & \colhead{log$g$} & \colhead{$\xi$} &
\colhead{[Fe/H]}\\
\colhead{} &  [K] & [cm s$^{-2}$] & [km s$^{-1}$] & \colhead{}}
\startdata
BD+19 2531 &    5100 &  4.5 & 0.6 & 0.18 \\
HD 26913   &    5600 &  4.5 & 0.6 & 0.05 \\
HD 38392   &    5250 &  4.5 & 1.0 & 0.04 \\
HD 41593   &    5350 &  4.5 & 0.8 & 0.10 \\
HD 60491   &    5300 &  4.5 & 1.0 & -0.06 \\
HD 61606   &    5200 &  4.5 & 0.8 & 0.03 \\
HD 64942   &    5800 &  4.4 & 0.7 & 0.14 \\
HD 81659   &    5700 &  4.4 & 0.6 & 0.10 \\
HD 88654   &    5600 &  4.5 & 1.0 & 0.25 \\
HD 98712   &    5450 &  4.5 & 1.7 & 0.04 \\
HD 131156A &    5550 &  4.5 & 0.7 & -0.17 \\
HD 165185  &    5900 &  4.4 & 0.7 & 0.07 \\
\enddata
\end{deluxetable}

\begin{deluxetable}{lccccccccc}
\tabletypesize{\scriptsize}
\tablecaption{Other Nearby Young Stars Radial Velocities}
\tablewidth{0pt}
\tablehead{
\colhead{Star} & \colhead{\# nights obs}&
\colhead{\# obs}&
\colhead{$v$sin$i$} &
\colhead{$\Delta$V$_{H}$} &
\colhead{$A_{pred}$} &
\colhead{$A_{obs}$}  &
\colhead{$\sigma_{\rm error}$} &\colhead{M$_{3\sigma}$} & 
\colhead{Notes}\\
\colhead{}
 & & & 
[km s$^{-1}$] &
[mag] &
[km s$^{-1}$] &
[km s$^{-1}$] &
[km s$^{-1}$] & [M$_{Jup}$] &   }
\startdata
HD 10008  & 2  &   5 & 1.0    & 0.05   & 0.01   & 0.06 & 0.05 & 1  & \\ 
HD 15013  & 2  &   4 & 9.3    & 0.05   & 0.11   & 0.07 & 0.07 & 1  & \\
HD 17925  & 6  &  14 & 4.2    & 0.04   & 0.04   & 0.04 & 0.04 & 1  & \\ 
HD 19668  & 2  &   5 & 7.0    & 0.07   & 0.12   & 0.03 & 0.05 & 1  & \\ 
HD 82443  & 6  &   7 & 5.2    & 0.09   & 0.11   & 0.08 & 0.06 & 2  & \\
HD 82558  & 8  &  23 & (28.0) & 0.14   & 0.86   & 0.34 & 0.09 & 8  & 1 \\
HD 85512  & 8  &  14 & \nodata& 0.04   & \nodata& 0.12 & 0.06 & \nodata  & \\ 
HD 91901  & 8  &  16 & 1.0    & 0.06   & 0.01   & 0.05 & 0.04 & 1  & \\
HD 92945  & 4  &  10 & 5.7    & 0.06   & 0.08   & 0.03 & 0.04 & 1  & \\
HD 96064  & 6  &  15 & 5.4    & 0.09   & 0.11   & 0.07 & 0.06 & 2  & \\ 
HD 102195 & 11 &  17 & 3.5    & 0.05   & 0.04   & 0.06 & 0.04 & 2  & \\ 
HD 108767B& 3  &  10 & 1.8    & \nodata& \nodata& 0.06 & 0.04 & \nodata  & \\ 
HD 110514 & 5  &   7 & 4.9    & 0.04   & 0.05   & 0.04 & 0.06 & 1  & \\ 
HD 113449 & 8  &  15 & 5.8    & 0.07   & 0.10   & 0.14 & 0.05 & 2  & \\
HD 124106 & 4  &  11 & 3.8    & 0.04   & 0.04   & 0.05 & 0.04 & 1  & \\ 
HD 130004 & 6  &  11 & 1.0    & 0.07   & 0.02   & 0.07 & 0.04 & 1  & \\ 
HD 130307 & 3  &   6 & 1.0    & 0.05   & 0.01   & 0.07 & 0.04 & 1  & \\ 
HD 140901 & 4  &   9 & 1.6    & 0.03   & 0.01   & 0.05 & 0.05 & 1  & \\ 
HD 149661 & 8  &  18 & 2.1    & 0.02   & 0.01   & 0.07 & 0.04 & 1  & \\ 
HD 166348 & 4  &   7 & \nodata& 0.06   & \nodata& 0.05 & 0.03 & \nodata  & \\ 
HD 180134 & 3  &   9 & 8.1    & 0.03   & 0.06   & 0.07 & 0.04 & 1  & \\
HD 184985 & 3  &   8 & \nodata& 0.02   & \nodata& 0.08 & 0.05 & \nodata  & \\ 
HD 186803 & 3  &   6 & \nodata& 0.04   & \nodata& 0.04 & 0.05 & \nodata  & \\ 
HD 187101 & 2  &   5 & 10.5   & 0.06   & 0.15   & 0.04 & 0.06 & 1  & \\
HIP 51317 & 8  &  13 &($<$3.0)& 0.08   & 0.06   & 0.08 & 0.10 & 2  & 2 \\
HIP 60661 & 12 &  16 & \nodata& 0.10   & \nodata& 0.13 & 0.05 & \nodata  & \\ 
HIP 67092 & 11 &  16 & \nodata& 0.11   & \nodata& 0.10 & 0.12 & \nodata  & \\ 
HIP 74995 & 8  &  13 &($<$2.1)& 0.11   & 0.05   & 0.14 & 0.16 & 2  & 2 \\
\enddata
\\
(1) $v$sin$i$ from \citet{kovari2004}
(2) $v$sin$i$ from \citet{delfosse1998}
\end{deluxetable}

\begin{deluxetable}{lcccc}
\tabletypesize{\scriptsize}
\tablecaption{Other Young Stars Stellar Parameters}
\tablewidth{0pt}
\tablehead{
\colhead{Star} & \colhead{$T_{\rm eff}$} & \colhead{log$g$} & \colhead{$\xi$} &
\colhead{[Fe/H]}\\
\colhead{} &  [K] & [cm s$^{-2}$] & [km s$^{-1}$] & \colhead{}}
\startdata
HD 10008  & 5350    & 4.5     & 0.8     & -0.08    \\
HD 15013  & 5350    & 4.5     & 0.9     & -0.11   \\
HD 17925  & 5600    & 4.5     & 0.7     & 0.08    \\
HD 19668  & 5500    & 4.5     & 0.9     & -0.04   \\
HD 82443  & 5300    & 4.5     & 0.8     & -0.12   \\
HD 82558  & \nodata & \nodata & \nodata & \nodata \\
HD 85512  & \nodata & \nodata & \nodata & \nodata \\
HD 91901  & 5200    & 4.6     & 0.8     & -0.04   \\
HD 92945  & 5200    & 4.5     & 0.9     & 0.03    \\
HD 96064  & 5400    & 4.5     & 0.7     & 0.02    \\
HD 102195 & 5300    & 4.5     & 0.6     & 0.03    \\
HD 108767B& 5100    & 4.6     & 0.8     & -0.01   \\
HD 110514 & 5400    & 4.5     & 0.8     & -0.03   \\
HD 113449 & 5350    & 4.5     & 0.5     & -0.03   \\
HD 124106 & 5150    & 4.5     & 0.6     & -0.17   \\
HD 130004 & 5250    & 4.6     & 0.9     & -0.24   \\
HD 130307 & 5100    & 4.5     & 0.9     & -0.21   \\
HD 140901 & 5600    & 4.5     & 0.8     & 0.01    \\
HD 149661 & 5250    & 4.5     & 0.9     & -0.01   \\
HD 166348 & \nodata & \nodata & \nodata & \nodata \\
HD 180134 & 6500    & 4.2     & 0.8     & -0.20   \\
HD 184985 & \nodata & \nodata & \nodata & \nodata \\
HD 186803 & \nodata & \nodata & \nodata & \nodata \\
HD 187101 & 6000    & 4.4     & 0.9     & 0.04    \\
HIP 51317 & \nodata & \nodata & \nodata & \nodata \\
HIP 60661 & \nodata & \nodata & \nodata & \nodata \\
HIP 67092 & \nodata & \nodata & \nodata & \nodata \\
HIP 74995 & \nodata & \nodata & \nodata & \nodata \\
\enddata
\end{deluxetable}

\end{document}